\documentclass{aa}

\usepackage[varg]{txfonts}
\usepackage{siunitx}
\usepackage[version=4]{mhchem}
\usepackage{multirow}
\usepackage{graphicx}
\usepackage[caption=false]{subfig}
\usepackage[strict]{changepage}
\usepackage{xcolor}
\usepackage{ulem}
\usepackage[colorlinks=true,citecolor=blue]{hyperref}
\usepackage{float}
\usepackage{amssymb}
\usepackage{amsmath}
\usepackage{lscape}
\usepackage{mathtools}
\usepackage{natbib}
\usepackage{pdflscape}
\usepackage{gensymb}

\defcitealias{zhou2023}{Paper~I}   

\DeclareSIUnit\jansky{Jy}

\newcolumntype{L}[1]{>{\raggedright\let\newline\\\arraybackslash\hspace{0pt}}m{#1}}
\newcolumntype{C}[1]{>{\centering\let\newline\\\arraybackslash\hspace{0pt}}m{#1}}
\newcolumntype{R}[1]{>{\raggedleft\let\newline\\\arraybackslash\hspace{0pt}}m{#1}}

\begin{document}

\title{The evolution of the mass--radius relation of expanding very young star clusters}
\author{Jian-Wen Zhou\inst{\ref{inst1}} 
\and Pavel Kroupa\inst{\ref{inst2}} \fnmsep \inst{\ref{inst3}}
\and
Wenjie Wu\inst{\ref{inst2}}
}
\institute{
Max-Planck-Institut f\"{u}r Radioastronomie, Auf dem H\"{u}gel 69, 53121 Bonn, Germany \label{inst1} \\
\email{jwzhou@mpifr-bonn.mpg.de}
\and
Helmholtz-Institut f{\"u}r Strahlen- und Kernphysik (HISKP), Universität Bonn, Nussallee 14–16, 53115 Bonn, Germany \label{inst2}\\
\email{pkroupa@uni-bonn.de}
\and
Charles University in Prague, Faculty of Mathematics and Physics, Astronomical Institute, V Hole{\v s}ovi{\v c}k{\'a}ch 2, CZ-180 00 Praha 8, Czech Republic \label{inst3}
}

\date{Accepted XXX. Received YYY; in original form ZZZ}

\abstract
{
The initial mass--radius relation of embedded star clusters is an essentially important boundary condition for understanding the evolution of embedded clusters in which stars form to their release into the galactic field via an open star cluster phase. The initial mass--radius relation of embedded clusters deduced by Marks \& Kroupa is significantly different from the relation suggested by Pfalzner et al. Here, direct N-body simulations are used to model the early expansion of embedded clusters after the expulsion of their residual gas. The observational data of radii of up to a few~Myr old clusters collated from various sources are found to be well fit by the N-body models, implying these observed very young clusters to most likely be in an expanding state. We show that the mass-radius relation of Pfalzner et al. reflects the expansion of embedded clusters following the initial mass-radius relation of Marks \& Kroupa. We also suggest that even the embedded clusters in ATLASGAL clumps with HII regions are probably already in expansion. All here collected clusters from different observations show a mass-size relation with a similar slope, which may indicate that all clusters were born with a profile resembling that of the Plummer phase-space distribution function.
}


\titlerunning{}
\authorrunning{J. W. Zhou, Pavel Kroupa, W. Wu}

\maketitle 

\section{Introduction}
The formation of stars within molecular clouds and the early stages of stellar evolution are active research topics. 
Most and perhaps all the observed stars were formed in embedded clusters \citep{Kroupa1995a-277, Kroupa1995b-277, Lada2003-41,Kroupa2005-576,Megeath2016-151, Dinnbier2022-660}. 
The central hub of a hub-filament system in molecular cloud could serve as the precursor of an embedded cluster. In hub-filament systems, converging flows are funneling matter into the hub through the filaments, such that cores embedded in the hub can prolong the accretion time for growing massive stars due to the sustained supply of matter \citep{Myers2009,Schneider2012,Peretto2013,Longmore2014-291,Motte2018-56,Vazquez2019-490,Kumar2020-642,Zhou2022-514}.
Ultimately the feedback energy from the proto- and pre-main-sequence stars expels the remaining gas, the embedded cluster expands. Therefore,
measurements of the internal dynamics of young clusters and star-forming regions are necessary to fully understand the process of
their formation and dynamical evolution. 

Increasing observational evidence suggests that early expansion plays a fundamental role in the dynamical evolution of young star clusters. A useful visualization of the expansion of real star clusters is provided by the collation in fig.~4 in \cite{Brandner2008-387}. Gaia DR2 and DR3 data have opened a new window into the internal kinematics of young star clusters and large star-forming
complexes. 
For case studies, based on Gaia DR2 and DR3 data and multi-object spectroscopy, the expansion of very young clusters in star-forming regions has been revealed in many works \citep{Wright2019-486,Cantat2019-626,Kuhn2020-899,Lim2020-899,Swiggum2021-917,Lim2022-163,Muzic2022-668,Das2023-948,Flaccomio2023-670,Jadhav2024arXiv}. Statistically, in \citet{Kuhn2019-870}, for 28 clusters
and associations with ages of about~$1-5$ Myr, proper motions from Gaia DR2 reveal that at least 75\% of these systems are expanding. \citet{Della2023arXiv} presented a comprehensive kinematic analysis of virtually all known young ($t <$ 300 Myr) Galactic clusters based on the improved astrometric quality of the Gaia DR3 data, finding that a remarkable fraction (up to 80\%) of clusters younger than $\approx30\,$Myr is currently experiencing significant expansion. \citet{Wright2023arXiv} presented a large-scale 3D kinematic study of $\approx2000$ spectroscopically confirmed young stars ($t <20\,$Myr) in 18 star clusters and OB associations (groups) from the combination of Gaia astrometry and Gaia-ESO Survey spectroscopy. They found that nearly all of the groups studied are in the process of expansion. 

\citet{Bastian2006-369} compared the luminosity profiles of young massive clusters, such as M82-F, NGC 1569-A, and NGC 1705-1, with N-body simulations of clusters that had experienced rapid gas expulsion, resulting in the loss of a substantial portion of their mass. They found a strong agreement between the observed profiles and the simulations. \citet{Goodwin2006-373} analyzed the differences between luminosity-based and dynamical mass estimates for young massive stellar clusters. Their results revealed significant discrepancies, which they attributed to the possibility that many young clusters are not in virial equilibrium due to undergoing violent relaxation following gas expulsion. Additionally, they noted that the increasing core radii observed in young clusters within the Large Magellanic Cloud and Small Magellanic Cloud can be well explained as an effect of rapid gas loss. 
In \citet{Kroupa2001-321,Banerjee2013-764,Banerjee2014-787,Banerjee2015-447},
direct N-body modeling of realistic star clusters with a monolithically formed structure and undergoing residual gas expulsion have consistently reproduced the characteristics of several well-observed very young star clusters, i.e. the Orion Nebula Cluster (ONC), R136 and NGC 3603 clusters, thereby shedding light on the birth conditions of massive star clusters. Similar simulations were also carried out in \citet{Banerjee2017-597} to explain how young massive star clusters attain their current shapes and sizes from an initially compact embedded morphology. They found that young massive clusters cannot expand to their current sizes, only considering two-body relaxation, stellar mass loss, and dynamical heating through initial binaries. They also confirmed that a substantial residual gas expulsion with $\approx$30\% star formation efficiency can adequately swell the newborn embedded clusters. 
The simulations of \citet{Banerjee2013-764,Banerjee2014-787,Banerjee2015-447,Banerjee2017-597} start from an initially compact configuration of the computed cluster with the initial mass-radius relation of embedded clusters deduced by \citet{Marks2012-543}, i.e. 
\begin{equation}
 \frac{r_{\rm h}}{{\rm pc}}=0.10_{-0.04}^{+0.07}\times\left(\frac{M_{\rm ecl}}{M_{\odot}}\right)^{0.13\pm0.04}\;,
 \label{rm}
\end{equation}
This relation was derived from the radii and densities of open star clusters  and the binding energy distributions of  the surviving binary stars in them as these constrain the birth, or initial, half-mass radii of the precursor embedded star clusters. 
Based on the universality of the early cluster expansion, in this work,
we collected various cluster samples to characterize the early cluster expansion using the numerical simulation recipes in \citet{Kroupa2001-321,Banerjee2013-764,Banerjee2014-787,Banerjee2015-447,Dinnbier2022-660}, especially to study the evolution of the initial mass-radius relation of embedded clusters.


\section{N-body simulations}

\subsection{Basic parameters}\label{basic}
The parameters for the simulation were summarized from previous work, i.e.
\citet{Kroupa2001-321,
Baumgardt2007-380,
Banerjee2012-746,
Banerjee2013-764,
Banerjee2014-787,
Banerjee2015-447,
Oh2015-805,
Oh2016-590,
Banerjee2017-597,
Brinkmann2017-600,
Oh2018-481,
Wang2019-484,
Pavlik2019-626,
Wang2020-491,
Dinnbier2022-660}.
The previous simulations have already demonstrated the effectiveness and rationality of these parameter settings (see below). The influence of different parameter settings on simulation results and the discussion of multi-dimensional parameter space can also be found in the aforementioned literature and their references.
In this work, we mainly utilize the mature simulation recipe to interpret observational data. With this contribution we are not aiming to provide all possible solutions as this would entail searching for solutions in a high-dimensional space (cluster mass, age, age spread, star formation efficiency, time of onset of gas expulsion, initial spatial configuration of the embedded cluster - spherical or not, fractal or not). This is beyond the scope of a reasonable project. Instead, we test whether the initial conditions previously and independently deduced from other data and work provide reasonable physical representations of the stellar populations and their spatial distribution. 

We computed five clusters with the masses in stars of 100 M$_{\odot}$, 300 M$_{\odot}$, 1000 M$_{\odot}$ and 3000 M$_{\odot}$ in the first 4 Myr. The time step is 0.125 Myr.
The initial density profile of all clusters is the Plummer profile \citep{Aarseth1974-37, HeggieHut2003, Kroupa2008-760}.
The half-mass radius $r_{h}$ of the cluster is given by the $r_{\rm h}-M_{\rm ecl}$ relation (equation.\ref{rm}).
All clusters are fully mass segregated ($S$=1), no fractalization, and in the state of virial equilibrium ($Q$=0.5). $S$ and $Q$ are the degree of mass segregation and the virial ratio of the cluster, respectively. More details can be found in \citet{Kupper2011-417} and the user manual for the \texttt{McLuster} code.
The initial segregated
state is detected for young clusters and star-froming clumps/clouds \citep{Littlefair2003-345,Chen2007-134,Portegies2010-48,Kirk2011-727,Getman2014-787,Lane2016-833,
Alfaro2018-478,Plunkett2018-615, Pavlik2019-626, Nony2021-645, Zhang2022-936,
Xu2024-270}, but the degree of mass segregation is not clear. 
In simulations of the very young massive clusters R136 and NGC 3603 with gas expulsion by \citet{Banerjee2013-764}, 
mass segregation does not influence
the results. In Appendix.\ref{A}, we compared $S$=1 (fully mass segregated) and $S$=0.5 (partly mass segregated) in Fig.\ref{p-S}, and found the results are similar. 
We also discussed settings with and without fractalization in Fig.\ref{p-D}, the results of both were also consistent.

Sub-virial initial conditions of very young clusters have also been studied \citep{Adams2006-641,Proszkow2009-185,Proszkow2009-697}. An embedded cluster forms from a gas clump with the timescale about 1 Myr \citep{Kroupa2005-576,Kroupa2008-760,Zhou2024arXiv240903234Z} such that the sub-virial bulk state cannot persist for longer than this while the individual protostar takes about 0.1 Myr to reach most of its mass \citep{Wuchterl2003-398,Duarte2013-558}. Protostars thus decouple from the hydrodynamical flow and become ballistic particles \citep{Stutz2016-590} that orbit within the evolving overall potential while also having close encounters with other protostars.  
At the time of greatest compactness, when the embedded cluster has reached the radius-mass relation of embedded clusters \citep{Marks2012-543}, the embedded cluster can thus be assumed to be close to virial equilibrium since most of the protostars have orbited a few times within the contracting gas clump. An initial fractal or other sub-clustering can be set up in models, but such structures may be unrealistic because the protostars emerge from the gas over the crossing time scale such that there is never a fractal initial state with all stars present because it washes out within the crossing time scale. Using the observational data on NGC 3603 and R 136, \cite{Banerjee2015-447} and \cite{Banerjee2018-424} show that any initial sub-clustering needs to be so compact that it becomes essentially equal to monolithically formed clusters.
In short, the virialised state at the time when gas expulsion occurs is a reasonable assumption for the idealised initial conditions applied here.

The IMFs of the clusters are chosen to be canonical \citep{Kroupa2001-322} with the most massive star following the $m_{\rm max}-M_{\rm ecl}$ relation  of \citet{Weidner2013-434} and \citet{Yan2023-670}.
All stars are initially in binaries, i.e. $f_{\rm b}$=1 \citep{Kroupa1995-277-1491,Kroupa1995-277-1507,Belloni2017-471}, $f_{\rm b}$ is the primordial binary fraction.
We assume the clusters to be at solar metallicity, i.e. $Z=0.02$ \citep{von2016ApJ...816...13V}. 
The clusters traverse circular orbits within the Galaxy, positioned at a Galactocentric distance of 8.5 kpc, moving at a speed of 220 km s$^{-1}$.

The models here assume the canonical Kroupa/Belloni binary fraction of 100\% at birth \citep{Kroupa1995-277-1491,Kroupa1995-277-1507,Belloni2017-471}. The binary-star distribution functions underlying this assumption are the theoretical values which idealised star formation tends to if stars were to form in isolation. In a real embedded cluster binary systems with a separation larger than the distances between the protostars will not form and, for example, young protostars in Orion are observed to have a companion fraction of only 44\% down to reasonably close separations ($\approx$20 AU, see \citealt{Tobin2022-925}). The canonical Kroupa/Belloni distribution functions are however consistent with this because the wide binaries are immediately disrupted even prior to the first integration step in the N-body model (see Fig.12 in \citealt{Kroupa2008-760}). As these wide binaries contain, in their sum, a negligible amount of binding energy compared to that of the embedded cluster, their disruption does not affect the further evolution noticeably. The  kinematical cooling of an embedded cluster similar to the ONC (Orion Nebula Cluster) through the disruption of the wide binary population has been measured for the first time by \citet{Kroupa1999-4}, but the effect is negligible concerning the dynamical evolution of the embedded cluster.
The disruption of the birth binary population to a distribution of binary systems that is in equilibrium with the embedded cluster takes about a cluster crossing time ($<1\,$Myr). For the birth density of the ONC, $\approx 10^5\,M_\odot/$pc$^3$, it is about 40~per cent (see Fig.10 in \citealt{Marks2011-417}), in good agreeement with the above-mentioned observational result in \citet{Tobin2022-925}.
Anyway, in Appendix.\ref{A}, we compared $f_{\rm b}$=1 and $f_{\rm b}$=0.44 in Fig.\ref{p-b}, and found the results are similar.

Following previous simulations,
we consider the essential dynamical effects of the gas-expulsion process by applying a diluting, spherically-symmetric
external gravitational potential to a model cluster as in \citet{Kroupa2001-321}. Specifically, we use the potential
of the spherically-symmetric, time-varying mass distribution
\begin{eqnarray}
M_g(t)=& M_g(0) & t \leq \tau_d,\nonumber\\
M_g(t)=& M_g(0)\exp{\left(-\frac{(t-\tau_d)}{\tau_g}\right)} & t > \tau_d.
\label{eq:mdecay}
\end{eqnarray}
Here, $M_g(t)$ is the total mass in gas which is spatially distributed with the Plummer density distribution \citep{Kroupa2008-760} and starts depleting with timescale $\tau_g$ after a delay of $\tau_d$.
The Plummer radius of the gas distribution is kept time-invariant \citep{Kroupa2001-321}.
This assumption is an approximate model of the effective gas reduction within the cluster in the situation that gas is blown out while new gas is also accreting into the cluster along filaments such that the gas mass ends up being reduced with time but the radius of the gas distribution remains roughly constant. As discussed in \citet{Urquhart2022-510},
the mass and radius distributions of the ATLASGAL clumps at different evolutionary stages are quite comparable.
We use an average velocity of the outflowing gas of $v_g\approx10$ km s$^{-1}$ which is the typical sound-speed in an HII region. This gives
$\tau_g=r_h(0)/v_g$,
where $r_h(0)$ is the initial half-mass radius of the cluster.
We note that comparable outflows are detected in the about 0.1~Myr old Treasure Chest cluster \citep{Smith2005-888}
and in the few-Myr old star-burst clusters in the Antannae galaxies \citep{Whitmore1999-1551,Zhang2001-727}.
As for the delay-time, we take the representative value of $\tau_d\approx0.6$ Myr
(\citealt{Kroupa2001-321}, this being about the life-time of the ultra-compact HII region), and 
assume a star formation efficiency SFE $\approx$ 0.33, i.e. 
$M_{g}(0)$ = 2$M_{\rm ecl}(0)$.

Embedded clusters form in clumps. More massive clumps can produce more massive clusters, leading to stronger feedback and larger gas expulsion velocity \citep{Dib2013-436}. 
There should be correlations between the feedback strength, the clump (or cluster) mass and the gas expulsion velocity ($v_g$). And low-mass clusters should have a slower gas expulsion process compared with high-mass clusters. As shown in \citet{Pang2021-912}, low-mass clusters tend to agree with the simulations of slow gas expulsion. This work primarily focuses on low-mass star clusters, whose mass is far below that of young massive clusters studied in \citet{Bastian2006-369,Goodwin2006-373,Banerjee2013-764,Banerjee2014-787,Banerjee2017-597}. Therefore, a gas expulsion velocity $v_g\approx10$ km s$^{-1}$ {\color{red} is} upper-limit.
\citet{Yang2022-658} conducted the largest outflow survey so far toward ATLASGAL clumps mainly generated low-mass embedded clusters, the upper-limit of the outflow velocity, $\approx10\,$km~s$^{-1}$, is clearly shown in their Fig.5.

In Appendix.\ref{A}, we compared  SFE = 0.33 and SFE = 0.2 in Fig.\ref{p-e}. When SFE = 0.2, $M_{g}(0)$ = 4$M_{\rm ecl}(0)$, which is  twice that at SFE = 0.33. Hence, we also doubled  the gas expulsion timescale $\tau_g$.
Moreover, a large amount of gas can also affect the physical state of the cluster, such as it not being in virial equilibrium but instead collapsing due to excessive gravitational binding. Therefore, we set $Q$=0.2 for SFE = 0.2. 
In Fig.\ref{p-e}, clusters with SFE = 0.2 expand faster than with SFE = 0.33.
A smaller SFE implies a greater amount of gas, which affects the gas expulsion process, resulting in differences in the results. Further discussion of the details is not the focus of this work. 
Anyway, in consideration of the measurement errors in cluster ages, simulations with SFE = 0.2 can still roughly explain the observations, as shown in the following text.
In this work, we assumed a star formation efficiency (SFE) of 0.33. This value has been widely used in previous simulations listed above and has been proven to be effective. Such a SFE is also consistent with those obtained from hydrodynamic calculations including self-regulation \citep{Machida2012-421,Bate2014-437} and as well with observations of embedded systems in the solar neighbourhood \citep{Lada2003-41,Gutermuth2009-184,Megeath2016-151}.
In \citet{Zhou2024-688}, by comparing the mass functions of the ATLASGAL clumps and the identified embedded clusters, we found that a star formation efficiency of $\approx$ 0.33 can be a typical value for the ATLASGAL clumps, also presented in Fig.\ref{compare}(b). 
In \citet{Zhou2024arXiv240809867Z}, assuming SFE = 0.33, it was shown that the total bolometric luminosity of synthetic embedded clusters can well fit the observed bolometric luminosity of ATLASGAL clumps
with HII regions. 
In \citet{Zhou2024arXiv240903234Z}, we directly calculated the SFE of ATLASGAL clumps with HII regions, and found a median value of $\approx$0.3.


\subsection{Procedure}
The \texttt{McLuster} program \citep{Kupper2011-417} was used to set the initial configurations. 
The dynamical evolution of the model clusters was computed using the state-of-the-art \texttt{PeTar} code \citep{Wang2020-497}. 
\texttt{PeTar} employs the well-tested analytical single and binary stellar evolution recipes (SSE/BSE)
\citep{Hurley2000-315,Hurley2002-329,Giacobbo2018-474,Tanikawa2020-495,Banerjee2020-639}.

\subsection{Evolution of cluster radii}
\begin{figure}
\centering
\includegraphics[width=0.48\textwidth]{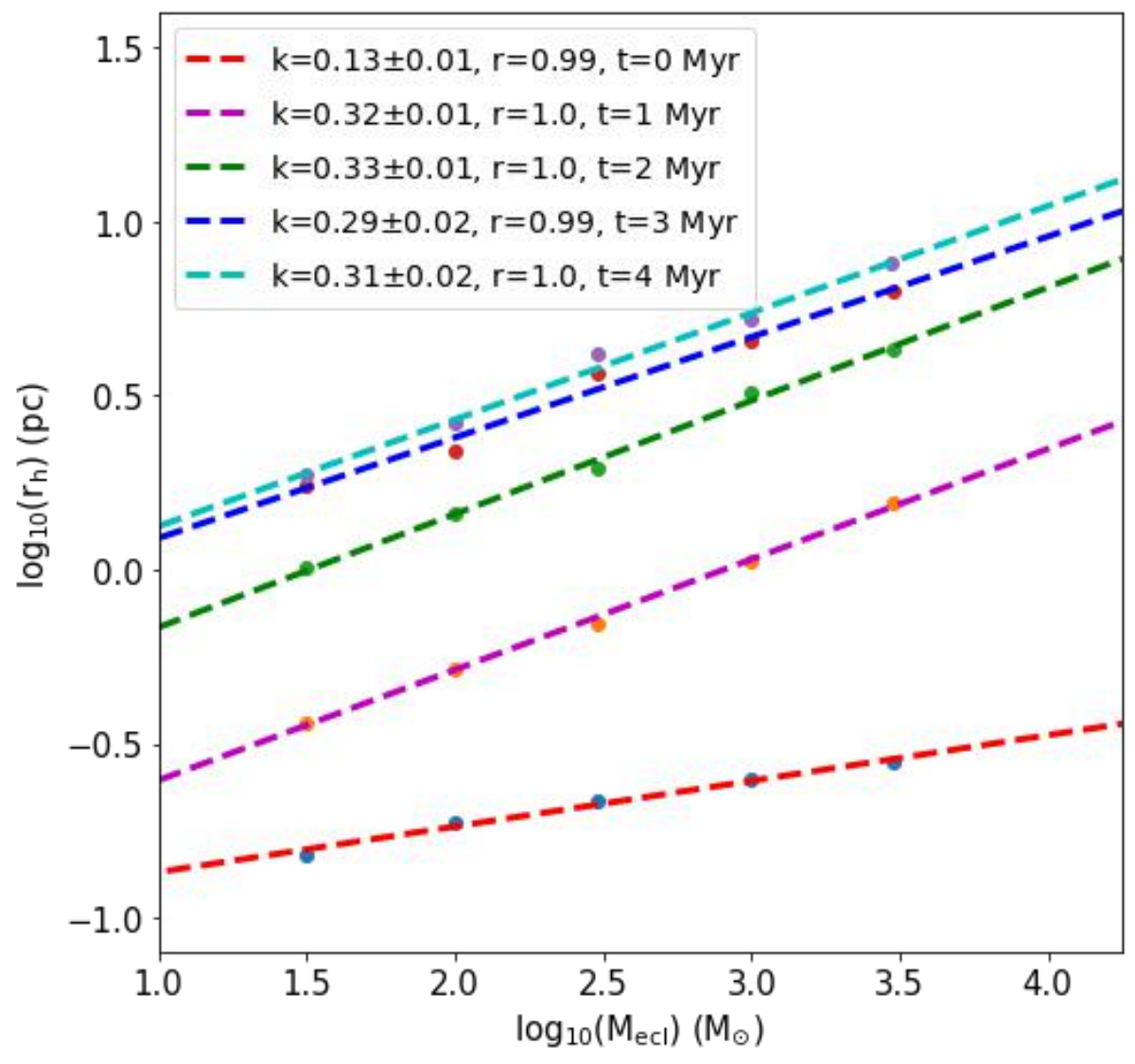}
\caption{
Fits to the expanding cluster by least-squares. The dots are from different time nodes, i.e. t = 0 Myr, 1 Myr, 2 Myr, 3 Myr and 4 Myr. 
``k'' is the slope of the linear fitting with the unit $\mathrm{pc/M_{\odot}}$.
``r'' represents the Pearson correlation coefficient.}
\label{fit}
\end{figure}

\begin{table}
\centering
\caption{Computed model parameters.
}
\label{tab1}
\begin{tabular}{ccccc}
\hline
Cluster	&	$M_{\rm ecl} (M_{\odot})$	&	$r_{\rm h} (pc)$	&	$m_{\rm max} (M_{\odot})$	&	$\tau_{\rm g} (Myr)$\\
1	&	30	&	0.16	&	5	&	0.012	\\
2	&	100	&	0.18	&	10	&	0.014	\\
3	&	300	&	0.21	&	17	&	0.016	\\
4	&	1000	&	0.25	&	30	&	0.019	\\
5	&	3000	&	0.28	&	43	&	0.022	\\
\hline
\label{para}
\end{tabular}
\end{table}

The computed model parameters are listed in Table.\ref{para}.
We studied the variations of the half-mass radius $r_{\rm h}$ (the radius within which is 0.5 of the mass in stars) and the full radius $r_{\rm f}$ 
(the radius within which is 0.9 of the mass in stars)
of the cluster in the first 4 Myr. The $r_{\rm h}-M_{\rm ecl}$ relation (equation.\ref{rm}) predicts the initial half-mass radius of the cluster at t = 0 Myr. In Fig.\ref{fit}, we fitted the $r_{\rm h}-M_{\rm ecl}$ relations at different time nodes, i.e. t = 1 Myr, 2 Myr, 3 Myr and 4 Myr. In the subsequent analysis, they are called the cluster's 1 Myr, 2 Myr, 3 Myr and 4 Myr expanding lines.
In the first 4 Myr the cluster mass in stars does not change significantly, we therefore assume stellar mass conservation, although the stars are allowed to evolve. 
To determine $r_{\rm h}(t)$ at a given mass of the embedded cluster in stars, $M_{\rm ecl}$, the distance of the star from the cluster center is documented within which half the mass of the model is found. At a time $t$, the data $r_{\rm h}(M_{\rm ecl}; t)$ are fitted by a linear relation using least-squares (Fig.~\ref{fit}).




\section{Cluster radii in observations}\label{benchmark}
In observations, different methods were used to estimate the size of the cluster, the measured radius can be the half-mass radius, $r_{\rm h}$ and the full radius, $r_{\rm f}$.
We can trace the evolution of these three radii at the same time in simulations, which can be the benchmarks to classify the observed radii of very young star clusters ($<$ 4 Myr).


\subsection{Half-mass radius}\label{rh}

\begin{figure}
\centering
\includegraphics[width=0.48\textwidth]{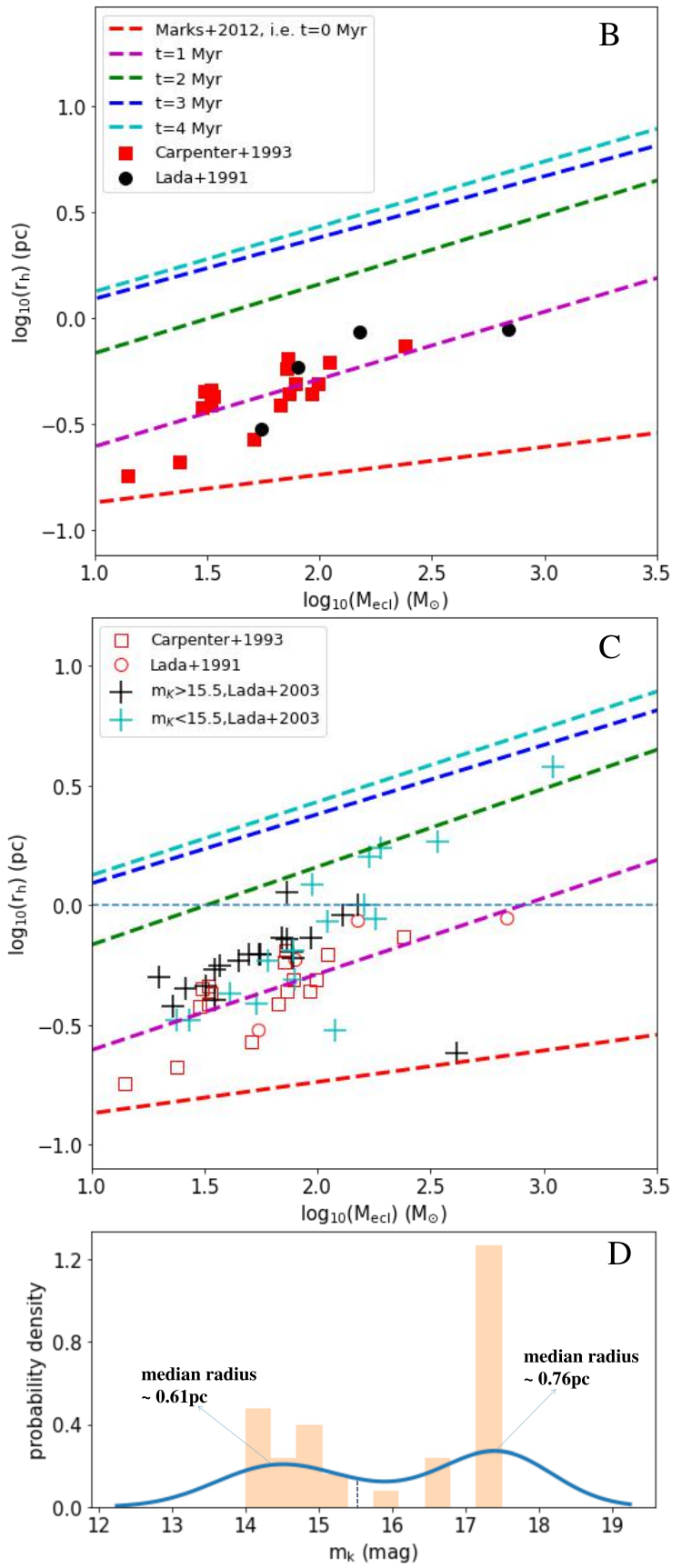}
\caption{(a) Dashed lines are the $r_{\rm h}-M_{\rm ecl}$ relations at different time nodes (t = 1 Myr, 2 Myr, 3 Myr and 4 Myr), i.e. the 1 Myr, 2 Myr, 3 Myr and 4 Myr expanding lines. Red squares and black dots represent the clusters from \citet{Carpenter1993-407} and \citet{Lada1991-371}, respectively; (b) "+" represent the clusters from \citet{Lada2003-41}, black and red "+" show the clusters selected by the magnitude limits $m_{\rm K}$<15.5 and $m_{\rm K}$>15.5, respectively.
Horizontal line marks the position of the radius $r=$ 1 pc;
(c) Distribution of the magnitude limits $m_{\rm K}$ of the clusters in \citet{Lada2003-41}. 
}
\label{bench}
\end{figure}

For the clusters in \citet{Lada1991-371}, they determined cluster boundaries using the magnitude limit of $m_{\rm K}$ < 13, and also required the lowest contour level to correspond to a surface density of 2 times the background density. In \citet{Carpenter1993-407}, the magnitude limit is $m_{\rm H}$ = 15.5. For the samples in these two papers, the boundaries of the clusters based on the identification criteria are probably not representative of the total extent of the cluster, thus, the measured radius should be smaller than the full radius $r_{\rm f}$ of the cluster. As shown in Fig.\ref{bench}(a), the samples can be well fitted by the 1 Myr expanding line. Considering the ages of these clusters are indeed around 1 Myr \citep{Weidner2013-434}, therefore, the simulations well explain the observed radii, and the measured radii in these observations are close to the half-mass radii of the clusters.

For the embedded cluster catalog in \citet{Lada2003-41}, as shown in Fig.\ref{bench}(b), the samples move systematically up, indicating older ages or larger radius measurements. 
The catalog gives the absolute magnitude limits $m_{\rm K}$ of the corresponding imaging observations. Fig.\ref{bench}(c) shows the distribution of the magnitude limits. If we take $m_{\rm K}$<15.5 and $m_{\rm K}$>15.5 as high and low standards of the magnitude limits, the median radii of the clusters in the two groups are $\approx$0.61 pc and 0.76 pc, respectively.
As expected, low standards indeed give slightly larger radius estimates due to the larger star counts. 
According to the selection criterion in \citet{Lada2003-41}, most of the clusters should have an age around 1 Myr. Actually, a part of the samples of \citet{Carpenter1993-407} and \citet{Lada1991-371} are in the catalog of \citet{Lada2003-41}. 
Except for those samples with significant deviations, the slightly systemic upshifting of most of the samples in \citet{Lada2003-41} in Fig.\ref{bench}(b) may be attributed to the low standards which cause slightly larger radius estimates. 
However, given that accurate age estimation of embedded clusters is always very challenging in observations, the differences may be just as likely caused by different ages of the cluster samples.

\subsection{Full radius}\label{rf}
\begin{figure}
\centering
\includegraphics[width=0.48\textwidth]{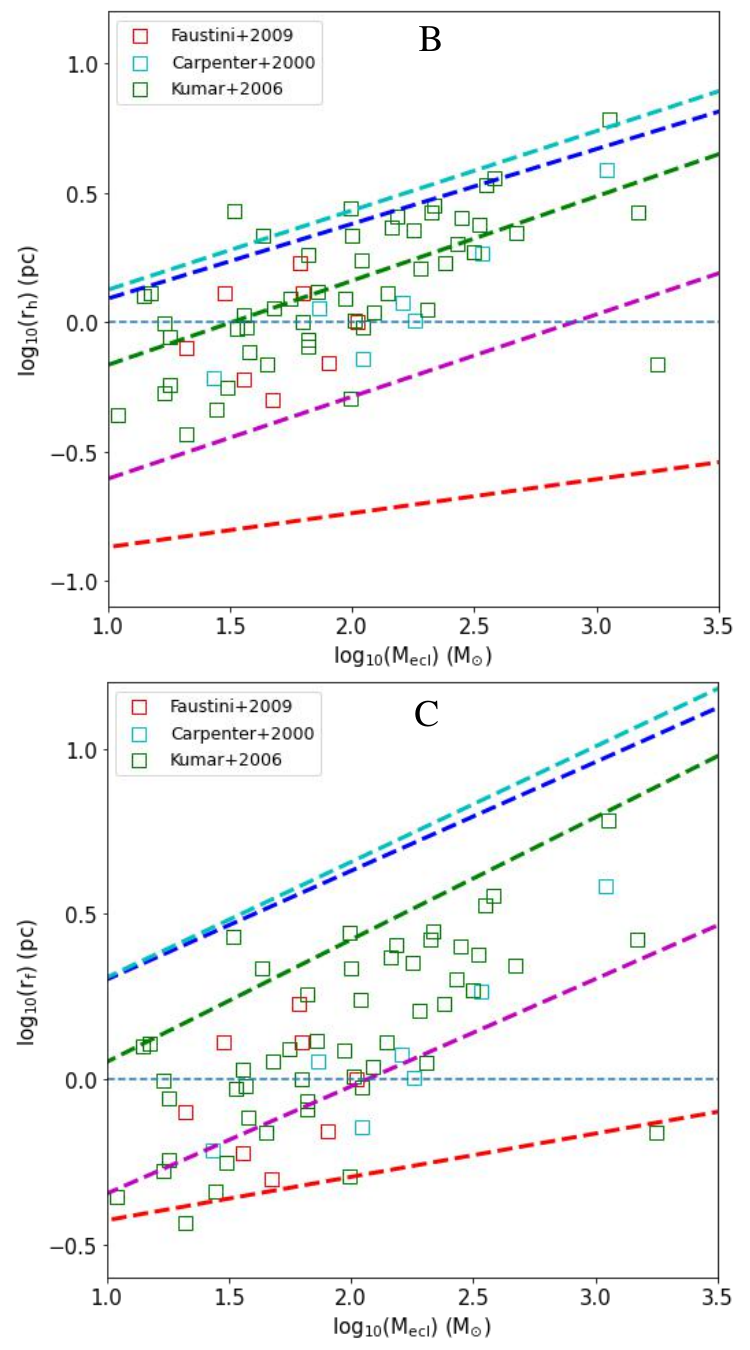}
\caption{The colored squares represent the clusters from \citet{Carpenter2000-120, Kumar2006-449, Faustini2009-503}. Horizontal line marks the position of the radius $r=$ 1 pc.
The dashed lines are color-coded with age as in Fig.\ref{fit} and Fig.\ref{bench}.
}
\label{full}
\end{figure}

In some observations, the cluster radius was measured by:
1. The surface density profile decreases until it reaches a constant value \citep{Faustini2009-503};
2. The density profile exceeds twice the standard deviation of the surface density in the surrounding field \citep{Carpenter2000-120, Kumar2006-449}. 
It seems these observations measured the full cluster radius.
In Fig.\ref{full}, we used the $r_{\rm h}-M_{\rm ecl}$ and $r_{\rm f}-M_{\rm ecl}$ relations to compare with the observations. The two relations give the age estimates of $\approx$2 Myr and $\approx$1 Myr, respectively. For the samples in \citet{Kumar2006-449}, they are the youngest stellar clusters associated with massive protostellar candidates, thus, their ages should be $\leq$1 Myr. Therefore, the $r_{\rm f}-M_{\rm ecl}$ relation gives a better fit than the $r_{\rm h}-M_{\rm ecl}$ relation, which means the above observations indeed roughly measured the full radii of the samples. The ages of the clusters in \citet{Carpenter2000-120} and \citet{Faustini2009-503} are $<$2 Myr and $\leq$1 Myr, respectively. All of them are well fitted in Fig.\ref{full}.

\subsection{Recent observations}

\begin{figure}
\centering
\includegraphics[width=0.48\textwidth]{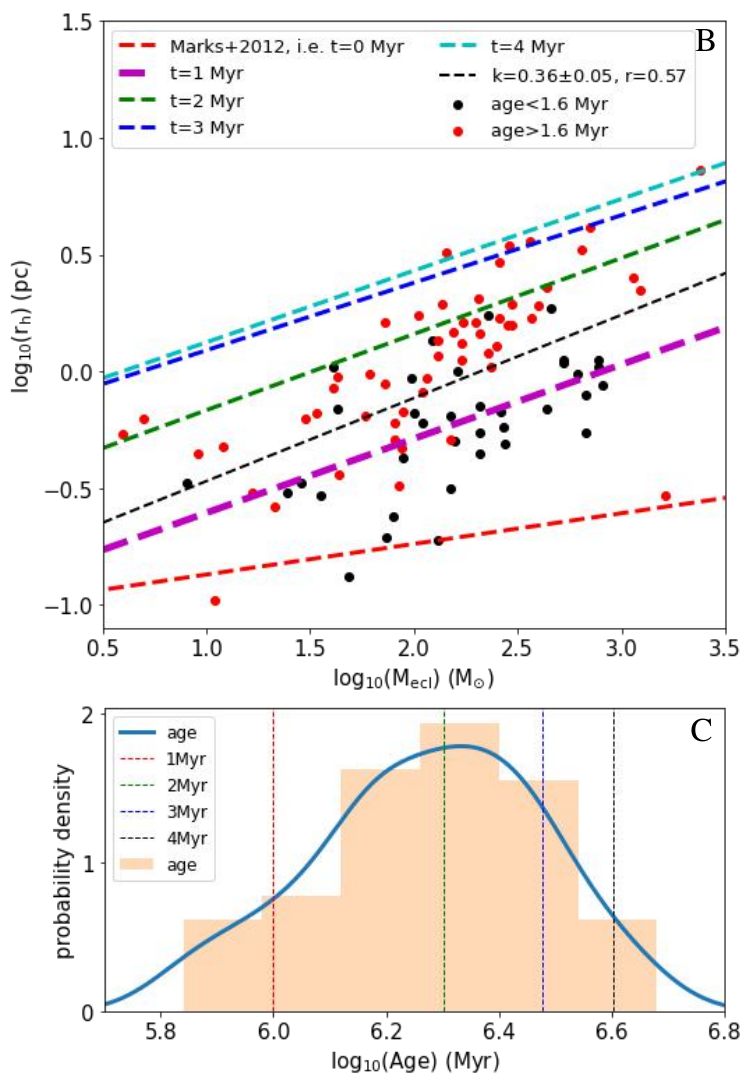}
\caption{
(a) Dashed lines are the $r_{\rm h}-M_{\rm ecl}$ relations at different time nodes. Red and black dots represent the subclusters from \citet{Kuhn2014-787} with the ages > 1.6 Myr and < 1.6 Myr, respectively. Dashed black line represents the fitted mass-size relation. ``k'' is the slope of the linear fitting with the unit $\mathrm{pc/M_{\odot}}$.
``r'' is the Pearson correlation coefficient; (b) Age distribution of the subclusters in Table.1 of \citet{Kuhn2015-812}.}
\label{kuhn}
\end{figure}

Readers may find the aforementioned sample dated and therefore question its reliability. Below, we further validate the above results using more recent observations.

The MYStIX (Massive Young Star-Forming Complex Study in Infrared and X-ray) project \citep{Feigelson2013-209} examined 20 nearby (d$<$3.6~kpc) massive star-forming regions (MSFRs) using a combination of archival {\it Chandra} X-ray imaging, {\it 2MASS+UKIDSS} near-IR (NIR), and {\it Spitzer} mid-IR (MIR) survey data.
The catalog of 31,784 young stars from the project includes both high-mass and low-mass stars, and disk-bearing and disk-free stars \citep{Broos2013-209}. 
Although the MYStIX samples are not ``complete'', the identification of young stars was performed in a uniform way for the different regions, enabling comparative analysis of stellar populations within these diverse regions.
Across 17 massive star-forming regions in the MYStIX project, 
\citet{Kuhn2014-787} identified 142 subclusters (their "subcluster" being synonymous to embedded cluster here) of young stars using finite mixture models. Their physical parameters were listed in \citet{Kuhn2015-812}, such as the size, the age and the total number of stars. \citet{Getman2014-787} provided age estimates for over 100 subclusters using the novel $Age_{JX}$ method, which encompass a wide age range between 0.5 and 5~Myr, showing that individual regions often have spatial segregation. 

The radii of subclusters in Table.1 of \citet{Kuhn2015-812} are four times the core radius based on their model and thought to be roughly the projected half-mass radii. As shown in Fig.\ref{kuhn}(a), 
the $r_{\rm h}-M_{\rm ecl}$ relations can indeed well fit the observations. However, the ages assigned to individual young stars may exhibit considerable uncertainty due to statistical errors in luminosities, uncertainties in dereddening, and the inherent variability in the X-ray luminosity-mass relation. Additionally, different evolutionary models might introduce systematic shifts in ages \citep{Getman2014-787,Kuhn2015-812}.
To further complicate the discussion, the identification of sub-clusters in \citet{Kuhn2014-787} is not based on the Plummer profile as used in the present simulations. 
Furthermore, the table only provides the total number of stars in each subcluster, and to derive mass, we assume an average stellar mass of 0.5 M$_{\odot}$. This may underestimate the mass of the subcluster. 
Using the samples from \citet{Lada2003-41}, in Fig.\ref{mk}, the actual correlation between the total number of stars and the mass of an embedded cluster systematically shifts upward compared with the 0.5 M$_{\odot}$ assumption.
Despite many uncertainties, the good fittings in Fig.\ref{kuhn}(a) remain impressive.
Actually, the expansion of subclusters has been confirmed in \citet{Kuhn2015-812} and \citet{Kuhn2019-870} from observations.

\section{Mass-size relation}\label{MR}
\begin{figure}
\centering
\includegraphics[width=0.48\textwidth]{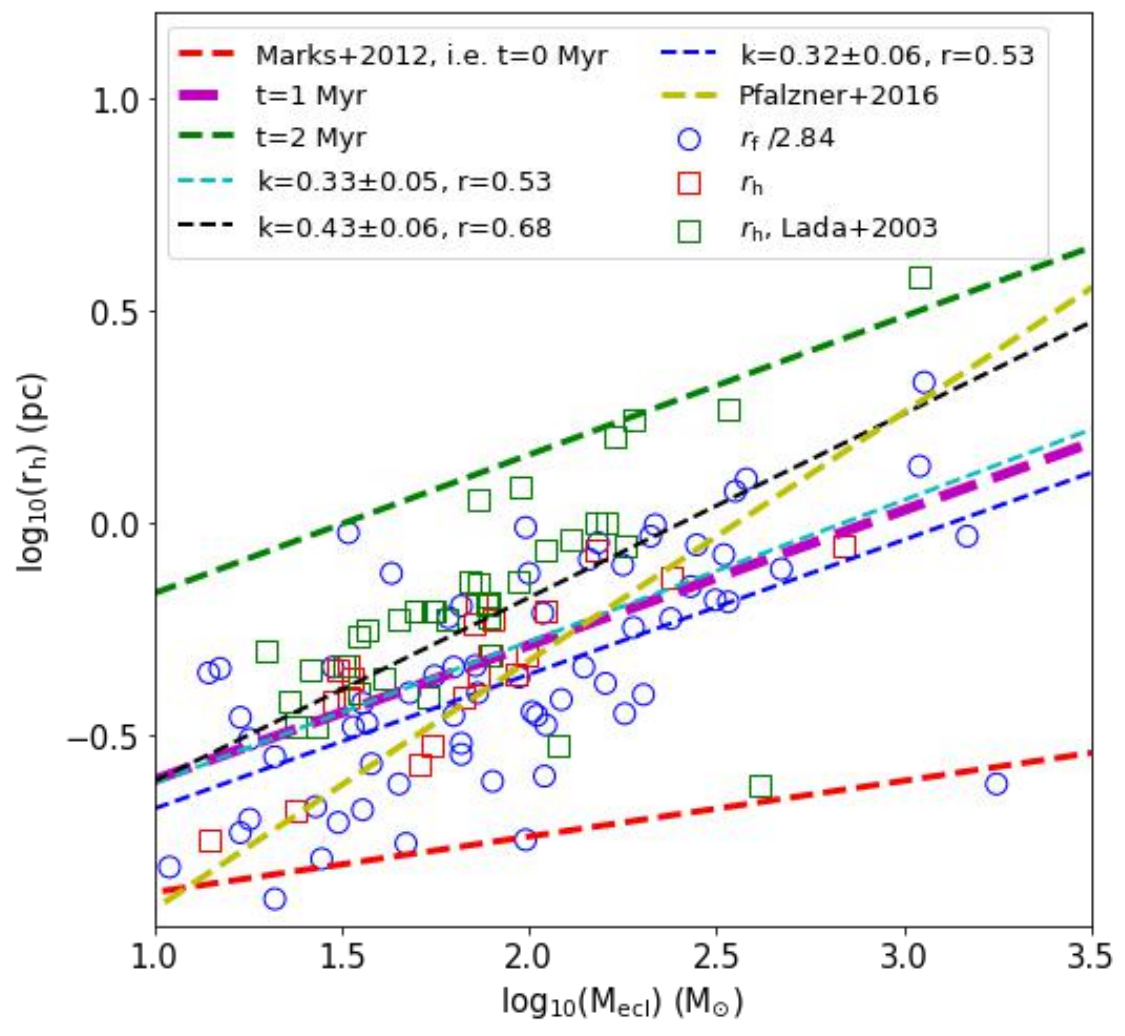}
\caption{For the samples in Fig.\ref{full}, their radii were divided by 2.84, and then merged with the samples in Fig.\ref{bench}. The radii of all samples were unified to the half-mass radius. Yellow dashed line is the mass-size relation of \citet{Pfalzner2016-586}. Dashed cyan, black and blue lines represent the mass-size relations of all samples, the samples in \citet{Lada1991-371, Carpenter1993-407, Lada2003-41} and the samples in \citet{Carpenter2000-120, Kumar2006-449, Faustini2009-503}, respectively.
``k'' is the slope of the fitting with the unit $\mathrm{pc/M_{\odot}}$. ``r'' represents the Pearson correlation coefficient.
}
\label{unify}
\end{figure}

Assuming all embedded clusters have the Plummer profile with the Plummer radius $r_{\rm pl}$, the mass within radius $r$ is
\begin{equation}
M(r)=M_{\rm ecl}\frac{(\frac{r}{r_{\rm pl}})^3}{[1+(\frac{r}{r_{\rm pl}})^2]^\frac{3}{2}}.
\end{equation}
The full radius (0.9 mass fraction) and the half-mass radius (0.5 mass fraction) satisfy $r_{\rm f} \approx 3.707 r_{\rm pl}$ and $r_{\rm h} \approx 1.305 r_{\rm pl}$, thus, $r_{\rm f} \approx 2.84 r_{\rm h}$ \citep{Kroupa2008-760}. 
For the samples discussed above, we can unify their radii to the half-mass radii. 
The radii of the samples in \citet{Lada1991-371, Carpenter1993-407, Lada2003-41} can be approximated to the half-mass radii. For the samples in \citet{Carpenter2000-120, Kumar2006-449, Faustini2009-503}, their radii need to be divided by 2.84.  
As shown in Fig.\ref{unify}, different samples can indeed merge together after the unification. 
We fitted the mass-size relation for all samples, which gives a slope $\approx$0.33 (dashed cyan line in Fig.\ref{unify}). If we fit the samples in \citet{Lada1991-371, Carpenter1993-407, Lada2003-41} and \citet{Carpenter2000-120, Kumar2006-449, Faustini2009-503} separately, the slopes are $\approx$0.43 (dashed black line in Fig.\ref{unify}) and $\approx$0.32 (dashed blue line in Fig.\ref{unify}), respectively. Only using the samples in \citet{Kuhn2015-812} gives a slope of $\approx$0.36 in Fig.\ref{kuhn}(a).

The equation (4) of
\citet{Pfalzner2016-586} has a steeper slope $\approx$0.58 (dashed yellow line in Fig.\ref{unify}). In \citet{Pfalzner2016-586}, they also found that all collected data in the K-band show a mass-size relation with a similar slope. Therefore, they argued that clusters might “grow” from low-mass clusters to their final mass without changing their size, and
their size would be determined by the extent of the clump. However, as shown in this work, the unique mass-radius relation of \citet{Pfalzner2016-586} actually reflects the expansion of embedded clusters. There is no self-similar mass-radius development during the evolution of embedded clusters, because the expanding lines in Fig.\ref{bench} are not parallel at the beginning. The real initial mass-radius relation of embedded clusters (equation.\ref{rm}) is significantly different from the relation of \citet{Pfalzner2016-586}. For any observable cluster, they are already in the state of expansion with different degrees. Thus, their current mass-radius relation can not reveal the initial conditions of cluster formation. 
In \citet{Pfalzner2016-586},
there is no cluster containing a thousand stars and having a half mass radius $<$ 0.4 pc in their samples, because all the samples are expanding and thus detach from the initial compact configuration predicted by the initial mass-radius relation of embedded clusters (equation.\ref{rm}).
As confirmed in the next section, even the embedded clusters in ATLASGAL clumps with HII regions (HII-clumps) are already in expansion.

In \citet{Pfalzner2016-586}, all collected samples show a mass-size relation with a similar slope, also shown in Fig.\ref{bench}, Fig.\ref{full}, Fig.\ref{kuhn}(a) and Fig.\ref{unify} in this work, although different methods were used to determine the radius and mass in different observations. 
As explained in \citet{Pfalzner2016-586},
assuming a unique mass-radius relation and a self-similar mass-radius development throughout, then different mass and radius measurements basically move the data points along the line of the mass-radius relation. In the discussion above, we have excluded this interpretation. Another simple explanation is that all clusters initially have the Plummer profile and then expand through the expulsion of residual gas as modelled in Sec.\ref{basic}. As done in Sec.\ref{MR}, different radii in different observations can be unified to the half-mass radii. This operation will not change the slope of the mass-size relation, which only shifts the line of the mass-size relation, as shown in \citet{Pfalzner2016-586} and this work, also see the discussion in \citet{Zhou2024-688}.


\section{Expansion of embedded clusters in HII-clumps}
\begin{figure}
\centering
\includegraphics[width=0.48\textwidth]{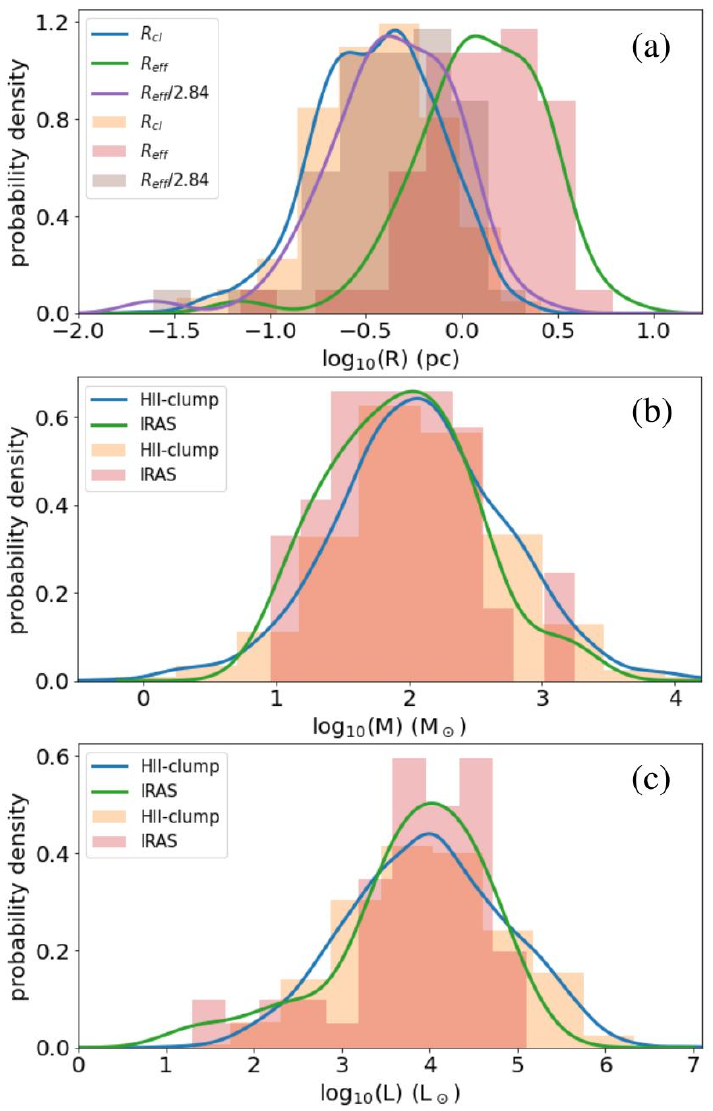}
\caption{Comparison of the radius, mass and bolometric luminosity of embedded clusters in HII-clumps of \citet{Urquhart2022-510} and the clusters (IRAS sources) in \citet{Kumar2006-449}.}
\label{compare}
\end{figure}
\begin{figure}
\centering
\includegraphics[width=0.48\textwidth]{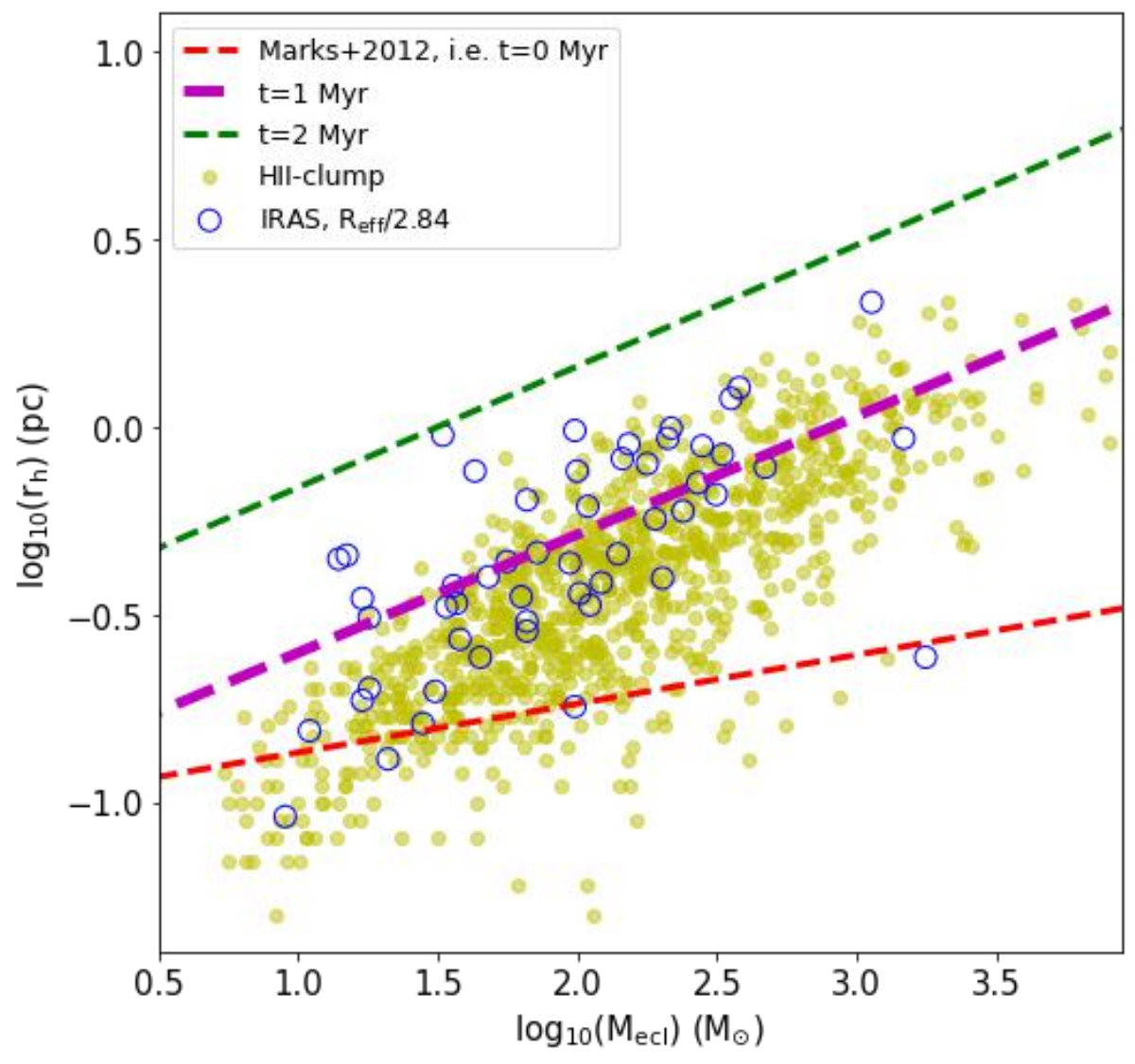}
\caption{Fitting the masses and sizes of embedded clusters in HII-clumps (yellow dots) using the expanding lines in Fig.\ref{fit} and Fig.\ref{bench}. Blue circles represent the clusters in \citet{Kumar2006-449}.}
\label{hii}
\end{figure}

For the sample in \citet{Kumar2006-449}, they are the youngest stellar clusters associated with massive protostellar candidates (IRAS sources), thus, they should be comparable with the embedded clusters in HII-clumps, as shown in Fig.\ref{compare} and Fig.\ref{hii}. The cluster radius in \citet{Kumar2006-449} is the effective radius $R_{\rm eff}$, which is approximated to the full radius of the cluster, as described in Sec.\ref{rf}. 
The radii of HII-clumps, $R_{\rm cl}$, in \citet{Urquhart2022-510} were determined from the number of pixels within the FWHM contour, i.e. above 50 per cent of the peak of the ATLASGAL dust continuum emission. As shown in Fig.\ref{compare}(a), $R_{\rm eff}$ is significantly larger than $R_{\rm cl}$. However, $R_{\rm eff}/2.84$ is comparable with $R_{\rm cl}$. $R_{\rm eff}/2.84$ is approximated to the half-mass radius of the cluster in \citet{Kumar2006-449}, as discussed in Sec.\ref{unify}. Therefore, $R_{\rm cl}$ measured in \citet{Urquhart2022-510} is close to the half-mass radius of the embedded cluster in a HII-clump. 
A HII-clump mass $M_{\rm cl}$ and its embedded star cluster mass $M_{\rm ecl}$ satisfy $M_{\rm ecl} = {\rm SFE} * M_{\rm cl}$, here ${\rm SFE}$ is the final star formation efficiency of the clump.
As discussed in Sec.\ref{basic}, we take SFE = 0.33.
In Fig.\ref{hii}, 
the masses and sizes of the embedded clusters in HII-clumps can be well fitted on the $r_{\rm h}-M_{\rm ecl}$ plane. Generally, the samples are between the 0--1 Myr expanding lines, consistent with their ages ($\leq$ 1 Myr). 
However, some samples are located under the initial 0 Myr line. This may be due to such clumps reaching a higher gas density during gas collapse than implied by the $r_{\rm h}-M_{\rm ecl}$ relation for embedded clusters, or the measurement of their radii might be biased or affected by resolution.
As shown in \citet{Urquhart2022-510}, evolutionary sequences still exist within HII-clumps. For the samples under the initial 0 Myr line, they may be far too young such that the inner cluster formation is not yet complete. 



\section{Conclusion}

We used standard N-body simulations as in previous works. The comparison between the previously deduced initial conditions of the simulations with the data used here has not been performed up to date, and this work constitutes a first step in an attempt to understand the data in terms of a realistic physical model that was developed previously on different data of young clusters and stellar populations. 

In this work, we traced the evolution of the $r_{\rm h}-M_{\rm ecl}$ and $r_{\rm f}-M_{\rm ecl}$ relations in the first 4 Myr by direct N-body simulations. These expanding lines at different time nodes as the benchmarks were used to classify the observed radii of very young star clusters. The good fit of the expanding lines to the samples compellingly suggests the observed very young clusters to be in expansion. After unifying the radii of all samples to the half-mass radii, we fitted the mass-size relations for different categories, and found that they are comparable with the mass-radius relation documented by \citet{Pfalzner2016-586}. Thus, the mass-radius relation of \citet{Pfalzner2016-586} actually reflects the expansion of embedded clusters starting with the initial mass-radius relation of \citet{Marks2012-543}. We also suggest that even the embedded clusters in ATLASGAL clumps with HII regions (HII-clumps) are already in expansion. 
All collected samples show a mass-size relation with a similar slope, which may indicate that embedded clusters are born with a profile resembling that of the Plummer phase-space density distribution function (e.g. \citealt{Aarseth1974-37, HeggieHut2003}). This is interesting, because the Plummer phase space density distribution function is the simplest analytical (two-parameter) distribution function that solves the time-independent collisionless Boltzmann equation (e.g., \citealt{Kroupa2008-760}), and because
the molecular cloud filaments in which stars have been found to have Plummer-like cross sections \citep{Andre2022-667},
and open star clusters are also well described by Plummer models \citep{Roeser2011-531,Roeser2019-627} as are globular star clusters \citep{Plummer1911-71}.



\bibliographystyle{aa} 
\bibliography{ref}

\appendix

\section{Appendix} \label{A}

In the simulations, all clusters are fully mass segregated ($S$=1) and no fractalization. We also assumed a star formation efficiency (SFE) of 0.33. However, some young clusters are observed to be highly substructured and do now show evidence of mass segregation. Moreover, the SFE also has a certain range of variation. Therefore, it is necessary to explore how varying these parameters might affect the cluster evolution and the mass-size relation. We compared fully mass segregated ($S$=1) and partly mass segregated ($S$=0.5) in Fig.\ref{p-S}, with and without fractalization in Fig.\ref{p-D}, and found the results are similar. 
Moreover, 100\% and 44\% initial binary fractions also give similar results, as shown in Fig.\ref{p-b}.
SFE=0.33 and SFE=0.2 in Fig.\ref{p-e} give different results, because
a smaller SFE implies a larger gas volume, which affects the gas expulsion process. 
However, given the measurement error of cluster age, simulations using SFE = 0.2 can still roughly explain the observations.

\begin{figure}
\centering
\includegraphics[width=0.48\textwidth]{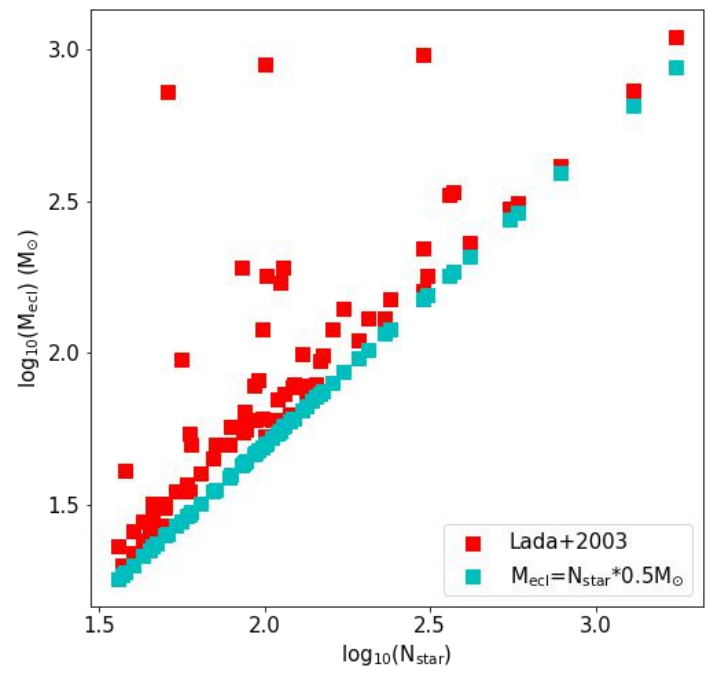}
\caption{Correlation between the total number of stars and the mass of embedded cluster.}
\label{mk}
\end{figure}

\begin{figure}
\centering
\includegraphics[width=0.48\textwidth]{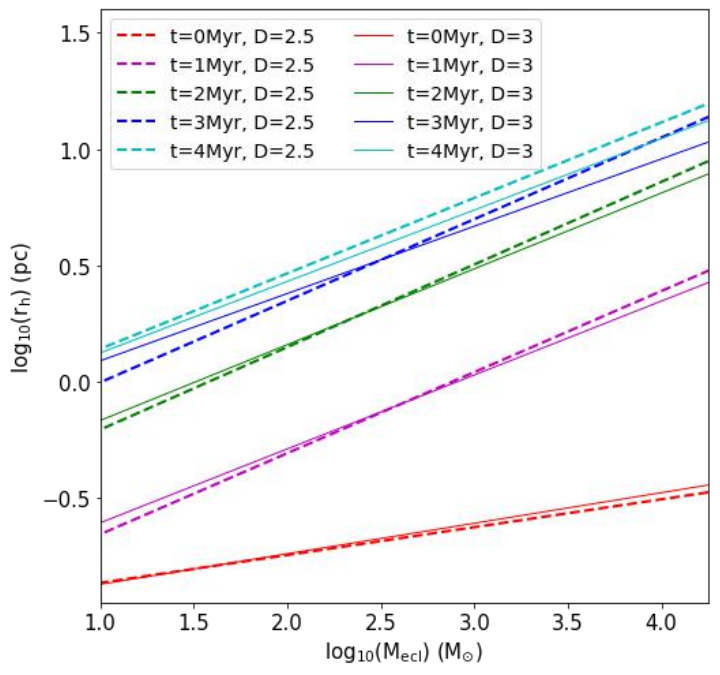}
\caption{Comparison of the expanding lines for clusters with and without fractalization. 
The settings for other parameters are the same as described in Sec.\ref{basic}.
We arranged a fractal distribution of stars within a sphere of constant average density, following a similar method as outlined in \citet{Goodwin2004-413}. The likelihood of a sub-box receiving a star can be represented by 2$^{D-3}$, where $D$ represents the fractal dimension. When $D$ is set to 3.0, there is no fractality observed since the probability becomes one. More details can be found in \citet{Kupper2011-417} and the user manual for the \texttt{McLuster} code.}
\label{p-D}
\end{figure}

\begin{figure}
\centering
\includegraphics[width=0.48\textwidth]{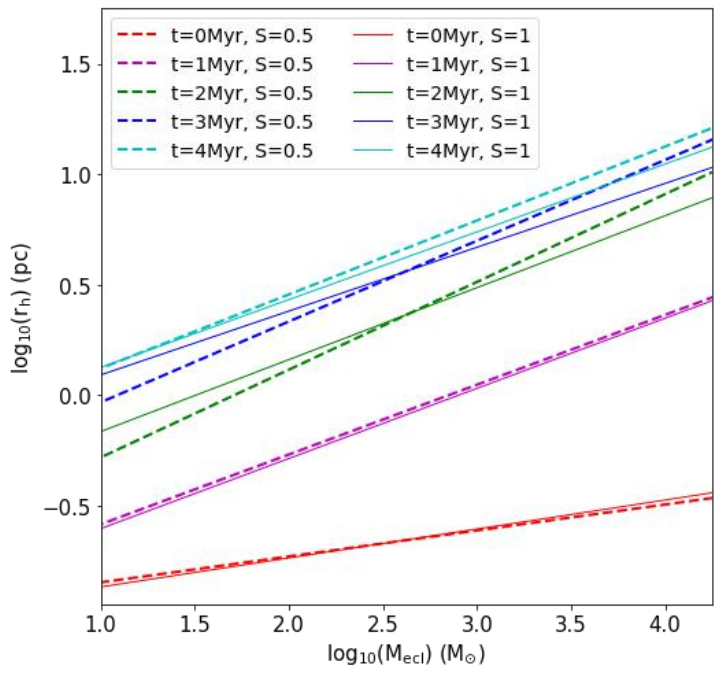}
\caption{Comparison of the expanding lines for clusters with fully mass-segregated ($S$=1) and partially mass-segregated ($S$=0.5) configurations. 
The settings for other parameters are the same as described in Sec.\ref{basic}.}
\label{p-S}
\end{figure}

\begin{figure}
\centering
\includegraphics[width=0.48\textwidth]{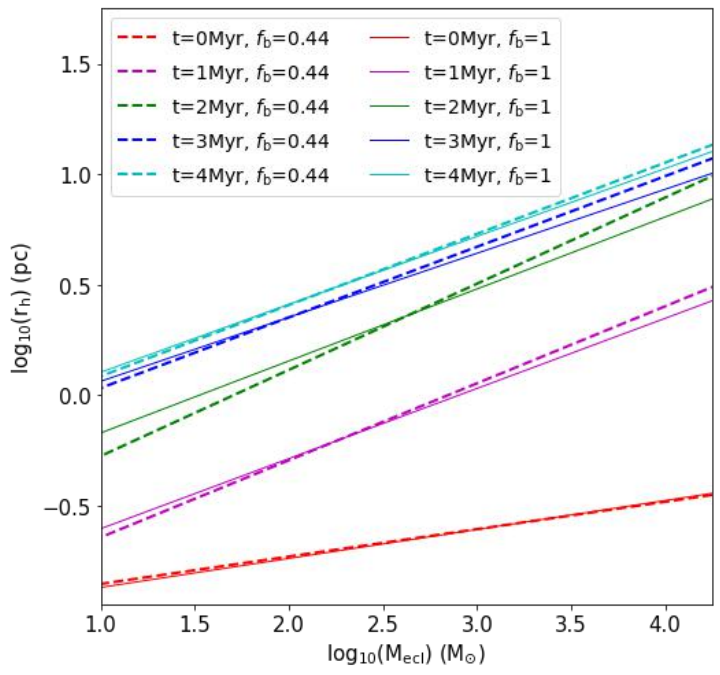}
\caption{Comparison of the expanding lines for clusters with 100\% ($f_{\rm b}$=1) and 44\% ($f_{\rm b}$=0.44) initial binary fractions. 
The settings for other parameters are the same as described in Sec.\ref{basic}.}
\label{p-b}
\end{figure}

\begin{figure}
\centering
\includegraphics[width=0.48\textwidth]{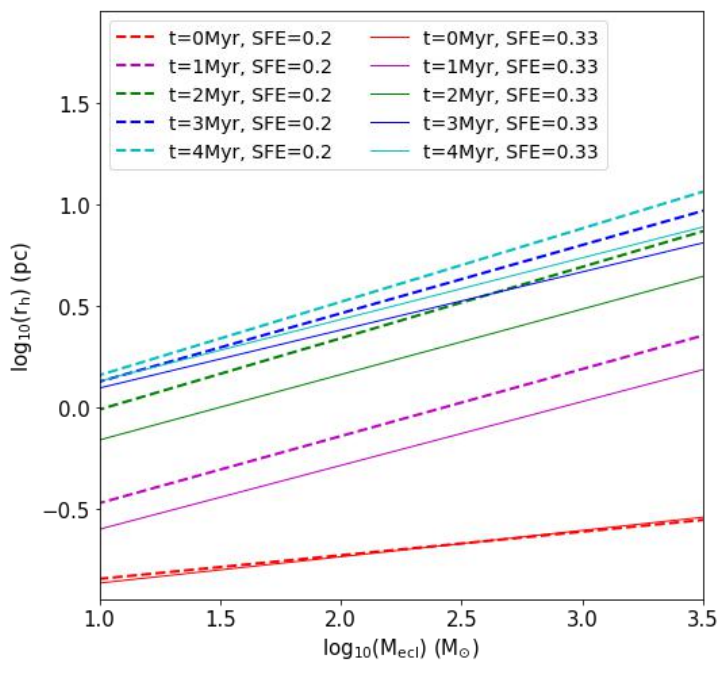}
\caption{Comparison of the expanding lines for clusters with SFE=0.33 and SFE=0.2. 
The settings for other parameters are the same as described in Sec.\ref{basic}.}
\label{p-e}
\end{figure}

\end{document}